\documentclass[10pt,onecolumn,twoside]{article}
\usepackage[dvips]{graphicx}
\usepackage[table,dvipsnames]{xcolor}
\usepackage{url}
\usepackage{graphicx}
\usepackage{algorithm,algorithmic}
\usepackage{bm}
\usepackage{amsmath,amssymb,array}
\usepackage{cite}
\usepackage{gensymb}
\usepackage{tikz}
\usepackage[breaklinks=true,colorlinks=false,bookmarks=false]{hyperref}
\usepackage{booktabs,siunitx}
\usepackage{multirow}
\usepackage{graphicx}
\usepackage{arydshln} 
\usepackage{mathtools} 
\usepackage{fullpage}

\def\argmin{\mathop{\mathrm{arg\,min}}} % Argument of a minimization
\def\prox{\mathrm{prox}} % Argument of a minimization
\def\defn{\,\coloneqq\,}
\def\xbm{{\bm{x}}}

\def\thetabm{{\bm{\theta}}}
\def\zerobm{{\bm{0}}}

\def\xbmhat{{\widehat{\bm{x}}}}

\def\R{\mathbb{R}}

\def\Tsf{{\mathsf{T}}}
\definecolor{brightgreen}{rgb}{0.4, 1.0, 0.0}
\definecolor{lightgreen}{rgb}{0.56, 0.93, 0.56}

\title{Joint Reconstruction and Calibration using\\Regularization by Denoising}

\author{Mingyang~Xie%
\thanks{Department of Computer Science \& Engineering, Washington University in St.~Louis, St.~Louis, MO 63130, USA.}
\hspace{0.05em},
Yu~Sun$^{\ast}$, Jiaming~Liu%
\thanks{Department of Electrical \& Systems Engineering, Washington University in St.~Louis, St.~Louis, MO 63130, USA.}
\hspace{0.05em},\\
Brendt~Wohlberg%
\thanks{Theoretical Division, Los Alamos National Laboratory,  Los Alamos, NM 87545 USA.}
\hspace{0.05em}, and Ulugbek~S.~Kamilov$^{\ast, \dagger}$}

\begin{document}
%\ninept
\date{}
\maketitle

\begin{abstract}
\noindent
\emph{Regularization by denoising (RED)} is a broadly applicable framework for solving inverse problems by using priors specified as denoisers. While RED has been shown to provide state-of-the-art performance in a number of applications, existing RED algorithms require  exact knowledge of the measurement operator characterizing the imaging system, limiting their applicability in problems where the measurement operator has parametric uncertainties. We propose a new method, called \emph{Calibrated RED (Cal-RED)}, that enables joint calibration of the measurement operator along with reconstruction of the unknown image. Cal-RED extends the traditional RED methodology to imaging problems that require the calibration of the measurement operator. We validate Cal-RED on the problem of image reconstruction in computerized tomography (CT) under perturbed projection angles. Our results corroborate the effectiveness of Cal-RED for joint calibration and reconstruction using pre-trained deep denoisers as image priors.
\end{abstract}

\section{Introduction}

The reconstruction of an unknown image $\bm{x}$ from a set of noisy measurements $\bm{y}$ is crucial in imaging inverse problems such as tomography~\cite{Wu.etal2020_simba, Sreehari.etal2016}, phase retrieval~\cite{Wu.etal2019a, Candes.etal2013}, and compressive sensing~\cite{Liu.etal2017, Metzler.etal2016}. When solving such ill-posed inverse problems, a widely used approach is to regularize the solution by using an image prior, such as the Tikhonov regularization~\cite{Ribes.Schmitt2008} or total variation regularization (TV)~\cite{Rudin.etal1992}.

Traditional image reconstruction methods assume accurate knowledge of the measurement operator characterizing the imaging system. In some cases, however, this operator is not known to sufficient accuracy, and can be modelled as depending on some parameters $\bm{\theta}$ that have to be calibrated to obtain an accurate characterization of the imaging system.
%This assumption overlooks imaging applications where the operator has some parametric uncertainty $\bm{\theta}$ that must be precisely calibrated so as to properly reconstruct $\bm{x}$. 
For example, calibration is essential in \emph{computerized tomography (CT)}~\cite{Kak.Slaney1988} when there is uncertainty in the projection angles used in data collection. In such settings, it is common to either manually calibrate the imaging system by using a known phantom~\cite{Cann.1988_ct_phantom_calibration, Wang.etal2007_ct_phantom, Miao.etal2012_ct_phantom, Zou.etal2013_ct_phantom} or embed a calibration step into the reconstruction method~\cite{Cheng.etal2018_taiwan_jiaotong, Lee.etal2017_phantomless, Malhotra.etal2016_cryo-em_india}.

There has been considerable recent interest in \emph{plug-and-play priors (PnP)}~\cite{Venkatakrishnan.etal2013} and \emph{regularization by denoising (RED)}~\cite{Romano.etal2017}, as frameworks for exploiting image denoisers as priors for image reconstruction. These methods achieve  state-of-the-art performance in a variety of imaging applications by integrating advanced denoisers that are not necessarily expressible in the form of a regularization function~\cite{Chan.etal2016, Teodoro.etal2019, Brifman.etal2016, Teodoro.etal2016, Zhang.etal2017a, Meinhardt.etal2017, Kamilov.etal2017, Sun.etal2019_online_pnp, Metzler.etal2018, Mataev.etal2019, Reehorst.Schniter2019}. However, current PnP/RED algorithms assume exact knowledge of the measurement operator, limiting their ability to solve problems involving parametric uncertainties in the data acquisition. 

% Sentence before revising
%  However, current PnP/RED algorithms assume exact knowledge of the measurement operator, limiting their ability to account for  parametric uncertainties in data acquisition. 

We propose a new \emph{calibrated RED (Cal-RED)} method that integrates an automatic parameter calibration procedure within the RED reconstruction. Each iteration of Cal-RED performs a $\bm{\theta}$-update step, followed by a traditional $\bm{x}$-update step, leading to the joint recovery of the unknown image $\bm{x}$ and parameters $\bm{\theta}$. One of the key benefits of Cal-RED is that it can leverage powerful \emph{deep denoisers}, such as DnCNN~\cite{zhang2017beyond}, for regularizing the inverse problem. We validate Cal-RED on CT by showing its ability to: \emph{(a)} reduce the \emph{root mean squared error (RMSE)} of the projection angles from $5\degree$ to $0.65\degree$; and \emph{(b)} improve the imaging \emph{signal to noise ratio (SNR)} from 10.16 dB (uncalibrated) to 21.13 dB (calibrated), with only a 0.95 dB gap relative to the oracle RED that knows the true angles. Note that while our focus in this letter is on RED, our calibration strategy is also fully compatible with PnP.

\section{Background}

\subsection{Inverse Problems in Imaging}

Consider the imaging problem specified by the linear model
\begin{equation}
	\bm{y}=\bm{H_{\theta}}\bm{x}+\bm{e},
\end{equation}
where $\bm{x} \in \mathbb{R}^n$ denotes the uknown image, $\bm{y} \in \mathbb{R}^{m}$ denotes the measurements, $\bm{H}_{\bm{\theta}} \in \mathbb{R}^{m \times n}$ is the measurement operator characterizing the response of the imaging system, $\bm{\theta} \in \mathbb{R}^\ell$ denotes the unknown parameters of the measurement operator, and $\bm{e} \in \mathbb{R}^{m}$ is the noise. When the true value of the parameters $\bm{\theta}$ is known, image reconstruction can be formulated as an optimization problem
\begin{equation}
	\xbmhat = \argmin\limits_{\bm{x} \in \R^n} f(\bm{x}) 
	\quad \text{with} \quad
	f(\bm{x})=g_\thetabm(\bm{x})+h(\bm{x}),
\end{equation}
where $g_\thetabm$ is the data-fidelity term that uses the measurement operator to ensure the consistency with the measurements and $h$ is the regularizer that imposes prior knowledge onto $\bm{x}$. For example, consider the smooth $\ell_2$-norm data-fidelity term
\begin{equation}
	g_\thetabm(\bm{x})= \frac{1}{2}||\bm{y}- \bm{H_{\theta}}\bm{x}||^{2}_{2} \;,
\label{formula_3}
\end{equation}
and the nonsmooth TV regularizer $h(\bm{x}) = \tau||\bm{D}\bm{x}||_{1}$, where $\tau>0$ is the regularization parameter and $\bm{D}$ is the image gradient~\cite{Rudin.etal1992}. \emph{Alternating direction method of multipliers (ADMM)}~\cite{Afonso.etal2010} and \emph{fast iterative shrinkage/thresholding algorithm (FISTA)}~\cite{Beck.Teboulle2009a} are two common optimization algorithms that address the nonsmoothness of $h$ in image reconstruction via the proximal operator
\begin{equation}
	\prox_{\tau h}(\bm{z}) \defn \argmin\limits_{\bm{x} \in\mathbb{R}^{n}}\left\{\frac{1}{2}||\bm{x}- \bm{z}||^{2}_{2} + \tau h(\bm{x}) \right\},
\label{formula_6}
\end{equation}
which can be interpreted as a \emph{maximum a prior probability (MAP)} estimator for AWGN denoising.

Deep learning has recently gained popularity for solving imaging inverse problems. An extensive review of deep learning in this context can be found in~\cite{McCann.etal2017, Lucas.etal2018, Knoll.etal2020}. Instead of explicitly defining an optimization problem, the traditional deep-learning approach is based on training an existing network architecture, such as UNet~\cite{Ronneberger.etal2015}, to invert the measurement operator by exploiting the natural redundancies in the imaging data~\cite{Han.etal2020_unet_conebeam, Sun.etal2018}. It is common to first bring the measurements to the image domain and train the network to map the corresponding low-quality images to their clean target versions via supervised learning~\cite{Han.etal2020_unet_conebeam, Sun.etal2018}. 

\subsection{Regularization by Denoising}
RED is a recently introduced image recovery framework \cite{Romano.etal2017, Reehorst.Schniter2019} that seeks an images $\bm{x}^*$ that satisfy
\begin{equation}
	\nabla g_\thetabm(\bm{x}^*) + \tau(\bm{x}^* - D_\sigma(\bm{x}^*)) = \zerobm \;,
\label{formula_4}
\end{equation}
where $\nabla g_\thetabm$ denotes the gradient of the data-fidelity term using the true calibration parameters $\thetabm$, $\tau$ is the regularization parameter, $D_\sigma$ is a denoiser for the input noise level $\sigma > 0$. RED thus computes an equilibrium point that balances the data fidelity against the fixed points of the denoiser. When the denoiser is locally homogeneous and has a symmetric Jacobian \cite{Reehorst.Schniter2019}, the term $\tau(\bm{x} - D_\sigma(\bm{x}))$ corresponds to the gradient of the regularization function
\begin{equation}
	h(\bm{x})= \frac{\tau}{2}\xbm^\Tsf(\bm{x} - D_{\sigma}(\bm{x})) \;.
% \label{formula_4}
\end{equation}
RED has been extensively applied in computational imaging, including in image restoration \cite{Mataev.etal2019}, phase retrieval \cite{Metzler.etal2018}, and tomographic imaging \cite{Wu.etal2020_simba}. Later works have developed  scalable variants of RED \cite{Sun.etal2019_online_pnp, Sun.etal2019b} and further replace the AWGN denoising with a general artifact-removal operator~\cite{Liu.etal2020}. 

\subsection{Calibrating the Measurement Operator}

The target-free calibration of the measurement operator has been an emerging topic in recent years. For CT scanners with slightly misaligned projection angles, existing calibration methods include entropy-based correction \cite{Donath.etal2006_ct_entrophy}, frequency-based correction \cite{Vo.etal2014_ct_frequency}, and a technique that combined gradient descent and multi-range testing \cite{Cheng.etal2018_taiwan_jiaotong}. A popular method to calibrate projection angles of Cryo-EM is called projection-matching \cite{Penczek_1994_projection_match, Baker.etal1996_proj_match}, which relies on a prior estimate of the density map of the unknown single particle. It seeks the  angle of every projection by pairing it with a clean template of the projection of the estimated density map from a known angle. Malhotraz {\it et al.} \cite{Malhotra.etal2016_cryo-em_india} propose a three-step 
pipeline for Cryo-EM to recover the projection angles and the target density map in sequence: (a) denoise the projections by the Principal Component Analysis technique, (b) estimate the angles through iterative coordinate gradient descent, and (c) reconstruct the density map by {\it filtered back-projection (FBP)}. The key contribution of this paper is the investigation of the calibration problem in the context of PnP/RED algorithms, which has not been done in the prior work.

\section{Proposed Method}

The proposed Cal-RED algorithm is summarized in Algorithm \ref{alg:cal-red}. It alternates between the updates of $\bm{\theta}$ and $\bm{x}$. We initialize $\bm{\theta}$ with their corresponding nominal values, $\hat{\bm{\theta}}$, which are assumed to be known from the design of the imaging system.
%Take a CT machine for example: $\hat{\bm{\theta}}$ should be the designed projection angles stated in its user manual, such as 0\degree, 2\degree, 4\degree,...180\degree, while $\bm{\theta}$ denotes its actual projection angles that are corrupted due to misalignment and are unknown to users, such as 0.01\degree, 2.01\degree, 3.99\degree,...17.99\degree.
We additionally initialize the image $\bm{x}$ with a direct inversion from measurements $\bm{y}$, which for CT reduces to the traditional FBP~\cite{Kak.Slaney1988}.

For the $\bm{\theta}$-optimization, as shown in lines 5-7 in Algorithm \ref{alg:cal-red}, we minimize the objective function
% formulate the following optimization problem with the $\ell_2$-norm regularizer:
\begin{equation} \label{eq:thetaupdate}
    \argmin\limits_{\bm{\theta}} \Big\{g_\thetabm(\bm{x}) + p(\bm{\theta})\Big\}\;\; \text{with} \;\;\; p(\bm{\theta}) = \frac{\tau_{\theta}}{2} ||\bm{\theta} - \hat{\bm{\theta}}||^2_{2}
\end{equation}
by performing the gradient descent
\begin{equation}
		\bm{\theta}^{k}\leftarrow \bm{\theta}^{k-1} - \gamma_{\theta} \left(\frac{\partial g_{\thetabm^{k-1}}(\bm{x}^{k-1})}{\partial \bm{\theta}^{k-1}} + \tau_{\theta}(\bm{\theta} - \hat{\bm{\theta}})\right) \;.
\end{equation}
In Equation \ref{eq:thetaupdate}, we consider the Tikhonov penalty as the prior for $\thetabm$. The partial derivative $\partial g_\thetabm(\bm{x}, \bm{\theta})/\partial \bm{\theta}$ is calculated by automatic differentiation \cite{autograd2018}. %using PyTorch \cite{torch2019}. 

% Note that the partial derivative $\partial g(\bm{x}, \bm{\theta})/\partial \bm{\theta}$ is calculated by finite difference approximation with respect to every $\theta_i$ in $\theta$,
% \begin{equation}
% 		\frac{\partial g(\bm{x}, \bm{\theta})}{\partial \bm{\theta}^{}_{i}} = \frac{g(\bm{x}, \bm{\theta} + \epsilon e_i) - g(\bm{x}, \bm{\theta})}{\epsilon} \;,
% \end{equation}
% where $\epsilon\rightarrow0$ is a very small positive scalar, and $e_i$ is a unit vector along the direction of $\bm{\theta}_i$. Such approximation allows our method to accommodate measurement operators that are difficult to differentiate.
We update $\bm{x}$ by following the RED framework, as shown in lines 9-11 of Algorithm \ref{alg:cal-red}. For the denoising prior $D_{\sigma}$, we choose a 17-layer DnCNN architecture in which all the convolution filters are of size $3\times3$ and every feature map has 64 channels. Note that Cal-RED does not modify the $\bm{x}$ update step of RED, which implies that the $\bm{\theta}$-update step of Cal-RED could also be integrated within PnP or other variants of RED, such as BC-RED~\cite{Sun.etal2019b}.

Cal-RED relies on Nesterov acceleration \cite{Nesterov2004} when computing both $\bm{x}$ and $\bm{\theta}$ updates by adding the acceleration sequence
\begin{equation}
q_k \leftarrow \frac{1}{2} \left(1+\sqrt{1+ 4q^2_{k-1}}\right)\;\; \text{with} \;\; q_1 = 1 \;..
\end{equation}
\noindent When $q_k = 1$ for all $k$, the algorithms reverts to the usual gradient method without acceleration. In Algorithm \ref{alg:cal-red}, the variables $\bm{s}$ and $\bm{u}$ store the intermediate values used for acceleration of $\bm{x}$ and $\bm{\theta}$, respectively.

% In a more general context, within each outer iteration, we can update $\bm{\theta}$ for $a$ iterations, and then update $\bm{x}$ for $b$ iterations. Here Cal-RED sets both $a$ and $b$ to $1$ and achieves good performance, as corroborated by our experiments. 

% In a more general context, within each outer iteration, we can update $\bm{\theta}$ for $a$ iterations, and then update $\bm{x}$ for $b$ iterations. While Cal-RED sets both $a$ and $b$ to $1$, let us consider an alternative strategy where both $a$ and $b$ are set to a large number. The main disadvantage of such a strategy is that when given a severely mismatched $\bm{\theta}$ (or $\bm{x}$), the iterative update of $\bm{x}$ (or $\bm{\theta}$) might collapse to a bad local minimum due to the mutual dependence between the two optimizations. Our strategy effectively overcomes such problem by instantly correcting both $\bm{x}$ and $\bm{\theta}$ from the previous iteration, which dramatically reduces the bias introduced by the other variable.

\begin{algorithm}[t!]
	\caption{Cal-RED}  
	\label{alg:cal-red}
	\begin{algorithmic}[1]

		\renewcommand{\algorithmicrequire}{\textbf{Input:}}
		\renewcommand{\algorithmicensure}{\textbf{Output:}}
		\REQUIRE $\bm{y} \in \mathbb{R}^{m}, \hat{\bm{\theta}} \in \mathbb{R}^{l}, \gamma_{x} > 0, \gamma_{\theta} > 0, \tau_{x} > 0, \tau_{\theta} > 0, \text{ and } \{q_k\}_{k\in \mathbb{N}}$
		\STATE $\bm{\theta} \leftarrow \hat{\bm{\theta}}$
		\STATE $\bm{x} \leftarrow \text{DirectInversion}_{\hat{\bm{\theta}}}(\bm{y})$
        % \STATE $ q_{0} \leftarrow 1$
		\FOR {$k = 1,2,... $}
		\STATE \textit{\# $\bm{\theta}$-update step}
		\STATE $G_{\theta}(\bm{x}^{k-1}) \leftarrow \frac{\partial g(\bm{s}^{k-1}, \bm{u}^{k-1})}{\partial \bm{u}^{k-1}} + \tau_{\theta} (\bm{u}^{k-1} - \hat{\bm{\theta}})$
        % \STATE $\mathsf{G}_{\theta}(\bm{x}^{k-1}) \leftarrow \nabla_{\theta} g(\bm{s}^{k-1}, \bm{u}^{k-1}) + \tau_{\theta} (\bm{u}^{k-1} - \hat{\bm{\theta}})$
		\STATE $\bm{\theta}^{k}\leftarrow \bm{u}^{k-1} - \gamma_{\theta} G_{\theta}(\bm{u}^{k-1}) $	
		\STATE $\bm{u}^k\leftarrow \bm{\theta}^{k} + \frac{q_{k-1} - 1}{q_k}(\bm{\theta}^{k} - \bm{\theta}^{k-1}) $
	
		\STATE \textit{\# $\bm{x}$-update step}
		\STATE $G_{x}(\bm{s}^{k-1}) \leftarrow\frac{\partial g(\bm{s}^{k-1}, \bm{\theta}^{k})}{\partial \bm{s}^{k-1}} + \tau_{x} (\bm{s}^{k-1} - D_{\sigma}(\bm{s}^{k-1}))$	

		\STATE $\bm{x}^{k} \leftarrow \bm{s}^{k-1} - \gamma_{x} G_{x}(\bm{s}^{k-1}) $

		\STATE $\bm{s}^k\leftarrow \bm{x}^{k} + \frac{q_{k-1} - 1}{q_k}(\bm{x}^{k} - \bm{x}^{k-1}) $

% 		\STATE \# Update the parameter that controls the acceleration
% 		\STATE $ q_k \leftarrow \frac{1}{2} \left(1+\sqrt{1+ 4q^2_{k-1}}\right) $

		\ENDFOR
	\end{algorithmic}
\end{algorithm}

\begin{figure}[t!]
	\centering
	\includegraphics[width=8.5cm]{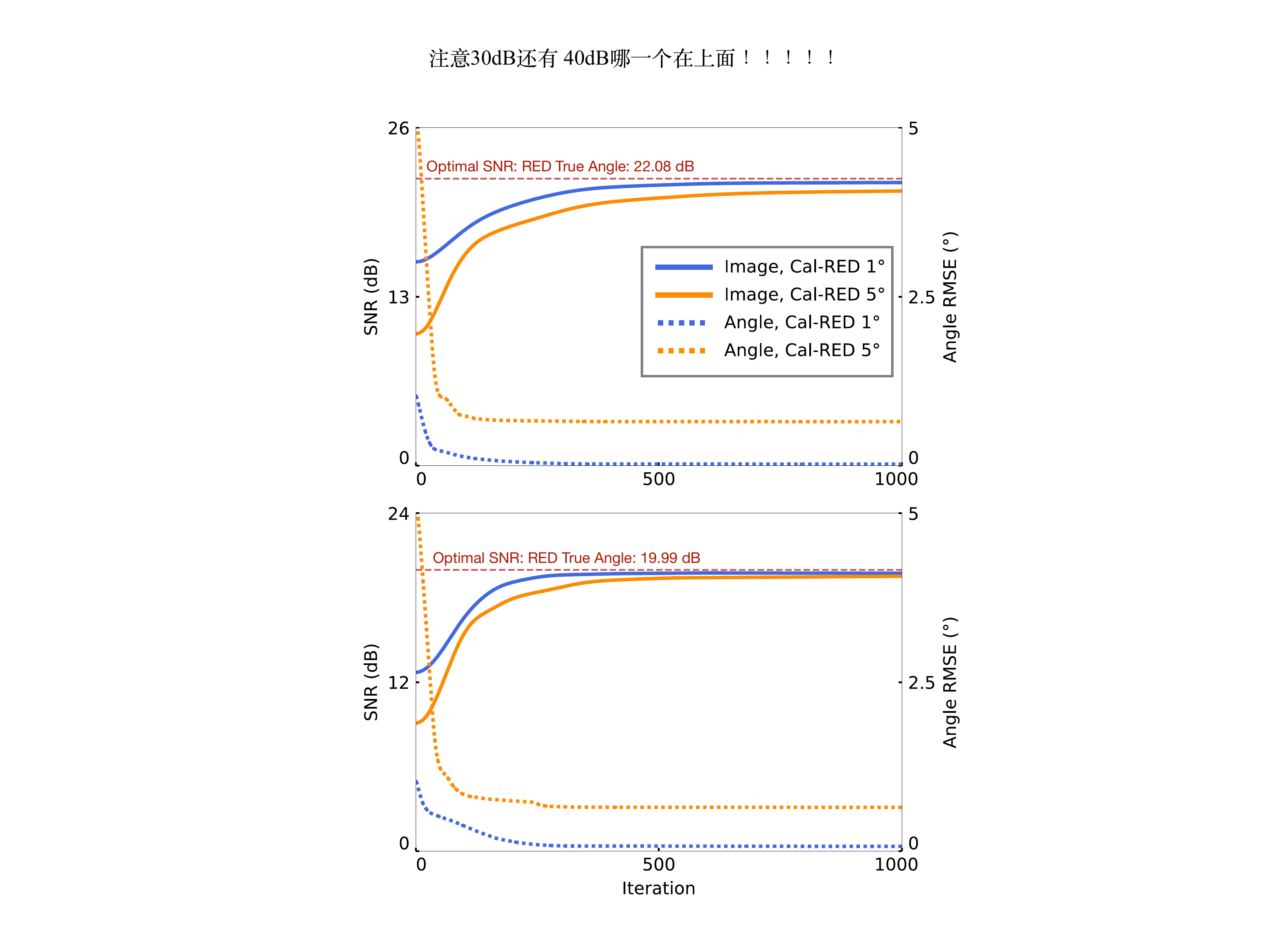}
	\caption{
		Illustration of the convergence of Cal-RED with projection angles corrupted by AWGN of SD 1\degree and 5\degree. The top figure corresponds to a sinogram input SNR of 40 dB, while the bottom figure corresponds to that of 30 dB. In each figure, the values of SNR and RMSE are plotted against the number of iterations using solid and dotted curves, respectively. The left axis represents the scale of SNR while the right axis represents that of RMSE. The optimal SNR values (obtained by RED using the true angles) is plotted as the horizontal dashed line for reference. Note the significant reduction of angular errors and the nearly optimal SNR results.
	}

	\label{snr_angle}
\end{figure}

\begin{figure*}[t!]
	\centering
\includegraphics[width=1.0\linewidth]{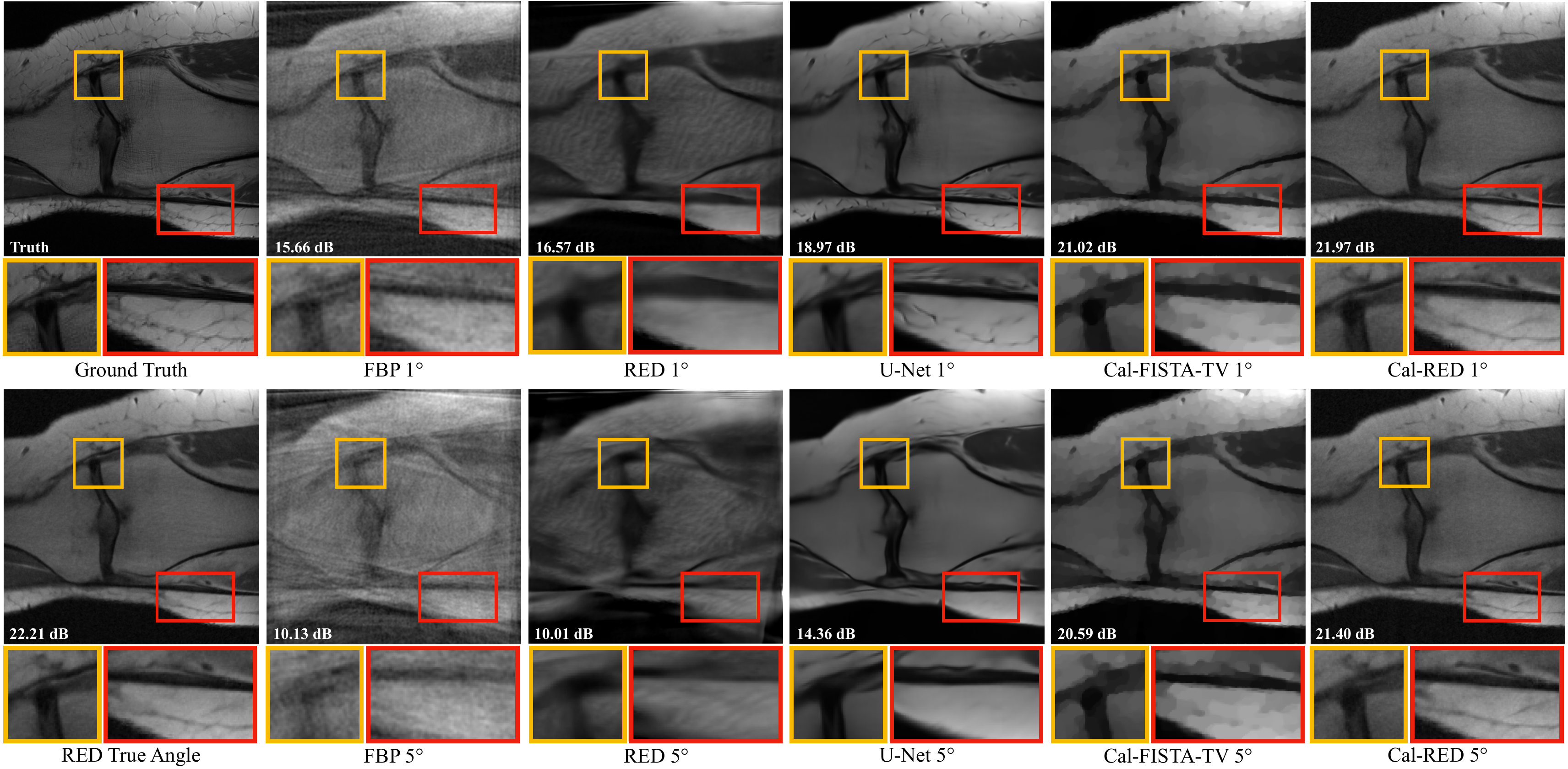}
	\caption{
		Recovery of a $320 \times 320$ knee image from a sinogram with noise corresponding to input SNR of 40 dB and projection angles corrupted by AWGN of SD 1\degree and 5\degree. Both the ground truth and the image reconstructed by oracle RED given the true angles are placed at the leftmost column for reference. The details are highlighted in the yellow and red boxes attached below each image.
	}
	\label{visual}
\end{figure*}
\section{Experiments}

In this section, we validate the performance of Cal-RED on simulated parallel-beam CT reconstruction, where the 2D target $\bm{x}$ is rotated around an axis to acquire projections from different angles and form a sinogram. We assume that the CT machine is designed to project from nominal angles $\hat{\bm{\theta}}$, an array of 90 projection angles that are evenly-distributed on a half circle. However, due to alignment errors, the CT machine actually projects from $\bm{\theta}$, another array of 90 angles that is assumed to be the result of corrupting $\hat{\bm{\theta}}$ by an unknown AWGN vector with standard deviation (SD) of 1\degree \;and 5\degree, respectively. In our experiments, we synthesize the sinograms by using the ground-truth $\bm{\theta}$ and further corrupt these sinograms by AWGN corresponding to the input SNR of 40 dB and 30 dB, respectively.

% In this section, we validate the performance of Cal-RED on simulated parallel-beam CT reconstruction, where the target 2-D object $\bm{x}$ is rotated around an axis to acquire projections from different angles to construct a sinogram. We assume that the CT machine is expected to project from an array of 90 projection angles, denoted by $\hat{\bm{\theta}}$, that are evenly-distributed on the half circle. Note that $\hat{\bm{\theta}}$ are known to the CT machine users. However, due to some sort of misalignment, the CT machine actually projects from another array of 90 angles, denoted by $\bm{\theta}$, which is assumed to be corrupted by an AWGN vector with standard deviation (SD) of 1\degree \;and 5\degree, respectively. Here, $\bm{\theta}$ is inaccessible during the image reconstruction.

We quantify the image quality by using SNR (dB)
\begin{equation}
\text{SNR}(\hat{\bm{x}}, \bm{x}) \triangleq 10 \log_{10} \frac{||\bm{x}||^2_{2}}{||\bm{x} - \hat{\bm{x}}||^2_{2}}
\end{equation}
and the angular error by RMSE 
\begin{equation}
\text{RMSE}(\hat{\bm{\theta}}, \bm{\theta}) \triangleq \sqrt{ \frac{\sum^{N}_{i=1}(\hat{\bm{\theta}}_i-\bm{\theta}_i)^2}{N}} \;.
\end{equation}
\noindent All numerical results are averaged over six medical knee images discretized to the size of $320\times320$ pixels. Note that the measurement operator $\bm{H}_\thetabm$ in this case corresponds to the Radon transform.

We compare Cal-RED with several baseline methods, including RED, U-Net, FISTA and the traditional least-square method (LSM). For both Cal-RED and RED, we train DnCNN denoisers for AWGN removal at three noise levels $\sigma \in \{5, 10, 15\}$ and pick the one that achieves the best performance for evaluation. We generate the training data for U-Net by filtered-back-projecting the sinogram using the mismatched angles $\hat{\bm{\theta}}$. We train an individual U-Net for each combination of $\{30\,\text{dB}, 40\,\text{dB}\} \times \{0\degree, 1\degree, 5\degree\}$. In order to better demonstrate the benefit of using deep denoising prior, we run FISTA with the TV regularizer for comparison. We also incorporate the $\bm{\theta}$-update step into both FISTA and LSM to construct the angle-calibrating FISTA (Cal-FISTA) and the angle-calibrating LSM (Cal-LSM), respectively.

Figure \ref{snr_angle} illustrates the convergence of Cal-RED in terms of both image SNR and angular RMSE. In the cases where $\bm{\theta}$ is severely mismatched, Cal-RED is able to reduce the angular RMSE by a factor of more than 7 --- from 5\degree to around 0.65\degree; moreover, Cal-RED can achieve a good reconstruction comparable to that of RED using the true $\bm{\theta}$ with respect to SNR. Table \ref{table_snr} and Table \ref{table_angle} summarize the numerical results under different combinations of input sinogram SNR and angular error. It is worth noting that Cal-RED and Cal-FISTA significantly outperform RED and FISTA, respectively, which highlights the effectiveness of the $\bm{\theta}$-update step. Despite being designed for the angle calibration purpose, when Cal-RED is applied to a problem without calibration errors, it achieves an SNR only 0.03 dB less than that of the original RED. When the input SNR is 30 dB, the angles recovered by Cal-RED and Cal-FISTA are far more accurate than those recovered by Cal-LSM, which demonstrates how denoising priors can also contribute to measurement operator calibration.

Figure \ref{visual} compares the visual results of FBP, U-Net, Cal-FISTA, Cal-RED, and RED. Both the reconstructions of FBP and uncalibrated RED suffer from obvious line artifacts, which demonstrates their vulnerability to a mismatched measurement operator. Although Cal-FISTA achieves an SNR close to that of Cal-RED, its reconstruction contains visible blocking artifacts. The U-Net removes most of the artifacts that are contained in the FBP reconstruction, but it also has an undesirable effect of oversmoothing small features. 
While the traditional deep learning alone fails to balance the preservation of features and the removal of artifacts, deep denoising priors like DnCNN help Cal-RED to reconstruct high-resolution images with minimal artifacts. For this particular image, when the projection angles are corrupted by AWGN of SD 5\degree, Cal-RED is able to achieve a remarkable SNR of \textcolor{black}{21.40 dB, while uncalibrated RED only achieves SNR of 10.01 dB}.
 
\begin{table}[tbp!]
	\centering
	\caption{
		Average SNRs of Reconstructed CT Images (dB)
	}
	\begin{tabular}{|c|c|c|c|c|c|c|}
		\hline
		\textbf{Input SNR}   &\multicolumn{3}{c|}{30 dB}& \multicolumn{3}{c|}{40 dB}  \\ \hline
% 		\textbf{sinogram}   & 30 dB  & 30 dB  & 30 dB  & 40 dB  & 40 dB  & 40 dB   \\ \hline
		\textbf{Angular Error}&0\degree&1\degree&5\degree &0\degree&1\degree&5\degree \\ \hline
		\textbf{FBP}        & 14.17  & 12.69  & 9.09    & 19.42  & 15.68  & 10.14   \\ \hline
		\textbf{U-Net}      & 19.84  & 18.01  & 14.49   & 22.01  & 18.58  & 14.55   \\ \hline
		\textbf{FISTA}      & 19.40  & 16.92  & 11.55   & 21.20  & 17.19  & 11.56   \\ \hline
		\textbf{RED}        & \textbf{\textcolor{ForestGreen}{19.99}}  & 15.46  & 10.06  & \textbf{\textcolor{ForestGreen}{22.08}}  & 15.84  &  10.16   \\ \hline
		\textbf{Cal-LSM}    & 7.80   & 5.07   & 4.18    & 16.73  & 14.88  & 13.34   \\ \hline

		\textbf{Cal-FISTA}  & 19.35      & 19.14  & 18.69   & 21.17      & 20.84  & 20.29   \\ \hline
		\textbf{Cal-RED}    & 19.96  & \textbf{\textcolor{ForestGreen}{19.74}}  & \textbf{\textcolor{ForestGreen}{19.52}}   & 22.07      & \textbf{\textcolor{ForestGreen}{21.78}}  & \textbf{\textcolor{ForestGreen}{21.13}}   \\ \hline
% 		\textbf{Cal-RED}    & 19.96  & \textcolor{brightgreen}{19.74}  & \textcolor{brightgreen}{19.52}   & 22.07      & \textcolor{brightgreen}{21.78}  & \textcolor{brightgreen}{21.13}   \\ \hline
	\end{tabular}
	\label{table_snr}
\end{table}

\begin{table}[tbp!]
	\centering
	\caption{
		Average RMSE of Calibrated CT Angles (\degree)
	}
	\begin{tabular}{|c|c|c|c|c|c|c|}
		\hline
		\textbf{Input SNR}   &\multicolumn{2}{c|}{30 dB}&\multicolumn{2}{c|}{40 dB}          \\ \hline
		\textbf{Angular Error}&1\degree&5\degree&1\degree&5\degree \\ \hline
		\textbf{Cal-RED}    & \textbf{\textcolor{ForestGreen}{0.073}}  & 0.650  & \textbf{\textcolor{ForestGreen}{0.024}}  & 0.651     \\ \hline
		\textbf{Cal-FISTA}  & 0.078  & \textbf{\textcolor{ForestGreen}{0.649}}  & 0.029  & \textbf{\textcolor{ForestGreen}{0.648}}     \\ \hline
		\textbf{Cal-LSM}    & 0.281  & 0.797  & 0.026  & 0.649     \\ \hline
	\end{tabular}
	\label{table_angle}
\end{table}

\section{Conclusion}

In this letter, we propose Cal-RED, a method that takes advantage of advanced denoising priors to jointly recover the measurement operator parameters and the target image. We validate its robustness to severely mismatched measurement operator parameters and noisy measurements via simulated CT reconstruction. Future work includes extending the proposed calibration strategy to other applications and image reconstruction frameworks.

\section*{Acknowledgements}

Research presented in this article was supported by the Laboratory Directed Research and Development program of Los Alamos National Laboratory under project number 20200061DR.

% \section{References}
\bibliographystyle{IEEEtran}

\begin{thebibliography}{10}
\providecommand{\url}[1]{#1}
\csname url@samestyle\endcsname
\providecommand{\newblock}{\relax}
\providecommand{\bibinfo}[2]{#2}
\providecommand{\BIBentrySTDinterwordspacing}{\spaceskip=0pt\relax}
\providecommand{\BIBentryALTinterwordstretchfactor}{4}
\providecommand{\BIBentryALTinterwordspacing}{\spaceskip=\fontdimen2\font plus
\BIBentryALTinterwordstretchfactor\fontdimen3\font minus
  \fontdimen4\font\relax}
\providecommand{\BIBforeignlanguage}[2]{{%
\expandafter\ifx\csname l@#1\endcsname\relax
\typeout{** WARNING: IEEEtran.bst: No hyphenation pattern has been}%
\typeout{** loaded for the language `#1'. Using the pattern for}%
\typeout{** the default language instead.}%
\else
\language=\csname l@#1\endcsname
\fi
#2}}
\providecommand{\BIBdecl}{\relax}
\BIBdecl

\bibitem{Wu.etal2020_simba}
Z.~{Wu}, Y.~{Sun}, A.~{Matlock}, J.~{Liu}, L.~{Tian}, and U.~S. {Kamilov},
  ``{SIMBA}: Scalable inversion in optical tomography using deep denoising
  priors,'' \emph{IEEE Journal of Selected Topics in Signal Process.}, vol.~14,
  no.~6, pp. 1163--1175, Mar. 2020.

\bibitem{Sreehari.etal2016}
S.~Sreehari, S.~V. Venkatakrishnan, B.~Wohlberg, G.~T. Buzzard, L.~F. Drummy,
  J.~P. Simmons, and C.~A. Bouman, ``Plug-and-play priors for bright field
  electron tomography and sparse interpolation,'' \emph{IEEE Trans. Comput.
  Imaging}, vol.~2, no.~4, pp. 408--423, Dec. 2016.

\bibitem{Wu.etal2019a}
Z.~Wu, Y.~Sun, J.~Liu, and U.~Kamilov, ``Online regularization by denoising
  with applications to phase retrieval,'' in \emph{IEEE Int. Conf. Computer
  Vision (ICCV) Workshops}, Oct. 2019.

\bibitem{Candes.etal2013}
E.~J. Cand{\`e}s, Y.~C. Eldar, T.~Strohmer, and V.~Voroninski, ``Phase
  retrieval via matrix completion,'' \emph{SIAM J. Imaging Sci.}, vol.~6,
  no.~1, Feb. 2013.

\bibitem{Liu.etal2017}
H.-Y. Liu, U.~S. Kamilov, D.~Liu, H.~Mansour, and P.~T. Boufounos,
  ``Compressive imaging with iterative forward models,'' in \emph{Proc. {IEEE}
  Int. Conf. Acoustics, Speech and Signal Process. ({ICASSP})}, New Orleans,
  LA, USA, Mar. 2017.

\bibitem{Metzler.etal2016}
C.~A. Metzler, A.~Maleki, and R.~G. Baraniuk, ``From denoising to compressed
  sensing,'' \emph{IEEE Trans. Inf. Theory}, vol.~62, no.~9, pp. 5117--5144,
  Sep. 2016.

\bibitem{Ribes.Schmitt2008}
A.~Rib{\'e}s and F.~Schmitt, ``Linear inverse problems in imaging,'' \emph{IEEE
  Signal Process. Mag.}, vol.~25, no.~4, pp. 84--99, Jul. 2008.

\bibitem{Rudin.etal1992}
L.~I. Rudin, S.~Osher, and E.~Fatemi, ``Nonlinear total variation based noise
  removal algorithms,'' \emph{Physica D}, vol.~60, no. 1--4, pp. 259--268, Nov.
  1992.

\bibitem{Kak.Slaney1988}
A.~C. Kak and M.~Slaney, \emph{Principles of Computerized Tomographic
  Imaging}.\hskip 1em plus 0.5em minus 0.4em\relax {IEEE}, 1988.

\bibitem{Cann.1988_ct_phantom_calibration}
C.~E. {Cann}, ``Quantitative {CT} for determination of bone mineral density: a
  review.'' \emph{Radiology}, vol. 166, no.~2, pp. 509--522, 1988.

\bibitem{Wang.etal2007_ct_phantom}
X.~Wang, J.~G. Mainprize, M.~P. Kempston, G.~E. Mawdsley, and M.~J. Yaffe,
  ``{Digital breast tomosynthesis geometry calibration},'' in \emph{Medical
  Imaging 2007: Physics of Medical Imaging}, vol. 6510, 2007.

\bibitem{Miao.etal2012_ct_phantom}
H.~Miao, X.~Wu, H.~Zhao, and H.~Liu, ``A phantom-based calibration method for
  digital {X}-ray tomosynthesis,'' \emph{Journal of X-ray Science and
  Technology}, vol.~20, pp. 17--29, Jan. 2012.

\bibitem{Zou.etal2013_ct_phantom}
J.~Zou, X.~Hu, H.~Lv, and X.~Hu, ``An investigation of calibration phantoms for
  {CT} scanners with tube voltage modulation,'' \emph{Int. Journal of
  Biomedical Imaging}, vol. 2013, Dec. 2013.

\bibitem{Cheng.etal2018_taiwan_jiaotong}
C.~Cheng, Y.~Ching, P.~Ko, and Y.~Hwu, ``Correction of center of rotation and
  projection angle in synchrotron {X}-ray computed tomography,''
  \emph{Scientific Reports}, vol.~8, Dec. 2018.

\bibitem{Lee.etal2017_phantomless}
D.~Lee, P.~Hoffmann, D.~Kopperdhal, and T.~Keaveny, ``Phantomless calibration
  of {CT} scans for measurement of bmd and bone strength-inter-operator
  reanalysis precision,'' \emph{Bone}, vol. 103, Aug. 2017.

\bibitem{Malhotra.etal2016_cryo-em_india}
E.~{Malhotra} and A.~{Rajwade}, ``Tomographic reconstruction from projections
  with unknown view angles exploiting moment-based relationships,'' in
  \emph{2016 IEEE Int. Conf. Image Process. (ICIP)}, 2016.

\bibitem{Venkatakrishnan.etal2013}
S.~V. Venkatakrishnan, C.~A. Bouman, and B.~Wohlberg, ``Plug-and-play priors
  for model based reconstruction,'' in \emph{Proc. IEEE Global Conf. Signal and
  Inf. Process. ({GlobalSIP})}, 2013.

\bibitem{Romano.etal2017}
Y.~Romano, M.~Elad, and P.~Milanfar, ``The little engine that could:
  {R}egularization by denoising ({RED}),'' \emph{SIAM J. Imaging Sci.},
  vol.~10, no.~4, pp. 1804--1844, 2017.

\bibitem{Chan.etal2016}
S.~H. Chan, X.~Wang, and O.~A. Elgendy, ``Plug-and-play {ADMM} for image
  restoration: Fixed-point convergence and applications,'' \emph{IEEE Trans.
  Comput. Imaging}, vol.~3, no.~1, pp. 84--98, Mar. 2017.

\bibitem{Teodoro.etal2019}
A.~M. Teodoro, J.~M. Bioucas-Dias, and M.~Figueiredo, ``A convergent image
  fusion algorithm using scene-adapted {G}aussian-mixture-based denoising,''
  \emph{IEEE Trans. Image Process.}, vol.~28, no.~1, pp. 451--463, Jan. 2019.

\bibitem{Brifman.etal2016}
A.~Brifman, Y.~Romano, and M.~Elad, ``Turning a denoiser into a super-resolver
  using plug and play priors,'' in \emph{Proc. {IEEE} Int. Conf. Image Proc.
  ({ICIP})}, Phoenix, AZ, USA, Sep. 2016.

\bibitem{Teodoro.etal2016}
A.~M. Teodoro, J.~M. Biocas-Dias, and M.~A.~T. Figueiredo, ``Image restoration
  and reconstruction using variable splitting and class-adapted image priors,''
  in \emph{Proc. {IEEE} Int. Conf. Image Proc. ({ICIP})}, Phoenix, AZ, USA,
  Sep. 2016.

\bibitem{Zhang.etal2017a}
K.~Zhang, W.~Zuo, S.~Gu, and L.~Zhang, ``Learning deep {CNN} denoiser prior for
  image restoration,'' in \emph{Proc. {IEEE} Conf. Computer Vision and Pattern
  Recognition ({CVPR})}, 2017.

\bibitem{Meinhardt.etal2017}
T.~Meinhardt, M.~Moeller, C.~Hazirbas, and D.~Cremers, ``Learning proximal
  operators: {U}sing denoising networks for regularizing inverse imaging
  problems,'' in \emph{Proc. IEEE Int. Conf. Computer Vision (ICCV)}, Venice,
  Italy, Oct. 2017.

\bibitem{Kamilov.etal2017}
U.~S. Kamilov, H.~Mansour, and B.~Wohlberg, ``A plug-and-play priors approach
  for solving nonlinear imaging inverse problems,'' \emph{IEEE Signal Process.
  Lett.}, vol.~24, no.~12, pp. 1872--1876, Dec. 2017.

\bibitem{Sun.etal2019_online_pnp}
Y.~{Sun}, B.~{Wohlberg}, and U.~S. {Kamilov}, ``An online plug-and-play
  algorithm for regularized image reconstruction,'' \emph{IEEE Trans. Comput.
  Imaging}, vol.~5, no.~3, pp. 395--408, 2019.

\bibitem{Metzler.etal2018}
C.~A. Metzler, P.~Schniter, A.~Veeraraghavan, and R.~G. Baraniuk, ``pr{D}eep:
  {R}obust phase retrieval with a flexible deep network,'' in \emph{Proc. 35th
  Int. Conf. Machine Learning ({ICML})}, 2018.

\bibitem{Mataev.etal2019}
G.~Mataev, M.~Elad, and P.~Milanfar, ``{DeepRED}: {D}eep image prior powered by
  {RED},'' in \emph{Proc. {IEEE} Int. Conf. Computer Vision Workshops
  ({ICCVW})}, Seoul, South Korea, Oct. 2019.

\bibitem{Reehorst.Schniter2019}
E.~T. Reehorst and P.~Schniter, ``Regularization by denoising: {C}larifications
  and new interpretations,'' \emph{IEEE Trans. Comput. Imaging}, vol.~5, no.~1,
  pp. 52--67, Mar. 2019.

\bibitem{zhang2017beyond}
K.~Zhang, W.~Zuo, Y.~Chen, D.~Meng, and L.~Zhang, ``Beyond a {Gaussian}
  denoiser: Residual learning of deep {CNN} for image denoising,'' \emph{IEEE
  Trans. Image Process.}, vol.~26, no.~7, pp. 3142--3155, Feb. 2017.

\bibitem{Afonso.etal2010}
M.~V. Afonso, J.~M.Bioucas-Dias, and M.~A.~T. Figueiredo, ``Fast image recovery
  using variable splitting and constrained optimization,'' \emph{IEEE Trans.
  Image Process.}, vol.~19, no.~9, pp. 2345--2356, Sep. 2010.

\bibitem{Beck.Teboulle2009a}
A.~Beck and M.~Teboulle, ``Fast gradient-based algorithm for constrained total
  variation image denoising and deblurring problems,'' \emph{IEEE Trans. Image
  Process.}, vol.~18, no.~11, pp. 2419--2434, November 2009.

\bibitem{McCann.etal2017}
M.~T. McCann, K.~H. Jin, and M.~Unser, ``Convolutional neural networks for
  inverse problems in imaging: A review,'' \emph{IEEE Signal Process. Mag.},
  vol.~34, no.~6, pp. 85--95, 2017.

\bibitem{Lucas.etal2018}
A.~Lucas, M.~Iliadis, R.~Molina, and A.~K. Katsaggelos, ``Using deep neural
  networks for inverse problems in imaging: {B}eyond analytical methods,''
  \emph{IEEE Signal Process. Mag.}, vol.~35, no.~1, pp. 20--36, Jan. 2018.

\bibitem{Knoll.etal2020}
F.~Knoll, K.~Hammernik, C.~Zhang, S.~Moeller, T.~Pock, D.~K. Sodickson, and
  M.~Akcakaya, ``Deep-learning methods for parallel magnetic resonance imaging
  reconstruction: {A} survey of the current approaches, trends, and issues,''
  \emph{IEEE Signal Process. Mag.}, vol.~37, no.~1, pp. 128--140, Jan. 2020.

\bibitem{Ronneberger.etal2015}
O.~Ronneberger, P.~Fischer, and T.~Brox, ``{U-Net}: {C}onvolutional networks
  for biomedical image segmentation,'' in \emph{Medical Image Computing and
  Computer-Assisted Intervention ({MICCAI})}, 2015.

\bibitem{Han.etal2020_unet_conebeam}
Y.~Han, J.~Kim, and J.~C. Ye, ``Differentiated backprojection domain deep
  learning for conebeam artifact removal,'' \emph{IEEE Trans. Medical Imaging},
  2020.

\bibitem{Sun.etal2018}
Y.~Sun, Z.~Xia, and U.~S. Kamilov, ``Efficient and accurate inversion of
  multiple scattering with deep learning,'' \emph{Opt. Express}, vol.~26,
  no.~11, pp. 14\,678--14\,688, May 2018.

\bibitem{Sun.etal2019b}
Y.~Sun, J.~Liu, and U.~S. Kamilov, ``Block coordinate regularization by
  denoising,'' in \emph{Advances in Neural Inf. Process. Systems 33},
  Vancouver, BC, Canada, Dec. 8-14, 2019.

\bibitem{Liu.etal2020}
J.~{Liu}, Y.~{Sun}, C.~{Eldeniz}, W.~{Gan}, H.~{An}, and U.~S. {Kamilov},
  ``{RARE}: Image reconstruction using deep priors learned without ground
  truth,'' \emph{IEEE J. Select. Topics Signal Process.}, 2020.

\bibitem{Donath.etal2006_ct_entrophy}
T.~Donath, F.~Beckmann, and A.~Schreyer, ``Automated determination of the
  center of rotation in tomography data,'' \emph{Journal of the Optical Society
  of America. A, Optics, Image science, and Vision}, vol.~23, pp. 1048--57,
  Jun. 2006.

\bibitem{Vo.etal2014_ct_frequency}
N.~T. Vo, M.~Drakopoulos, R.~Atwood, and C.~Reinhard, ``Reliable method for
  calculating the center of rotation in parallel-beam tomography,''
  \emph{Optics Express}, vol.~22, pp. 19\,078--19\,086, Jul. 2014.

\bibitem{Penczek_1994_projection_match}
P.~A. Penczek, R.~A. Grassucci, and J.~Frank, ``The ribosome at improved
  resolution: New techniques for merging and orientation refinement in 3{D}
  cryo-electron microscopy of biological particles,'' \emph{Ultramicroscopy},
  vol.~53, no.~3, pp. 251 -- 270, 1994.

\bibitem{Baker.etal1996_proj_match}
T.~S. Baker and R.~H. Cheng, ``A model-based approach for determining
  orientations of biological macromolecules imaged by cryoelectron
  microscopy,'' \emph{Journal of Structural Biology}, vol. 116, no.~1, pp. 120
  -- 130, 1996.

\bibitem{autograd2018}
A.~Baydin, B.~Pearlmutter, A.~Radul, and J.~Siskind, ``Automatic
  differentiation in machine learning: A survey,'' \emph{Journal of Machine
  Learning Research}, vol.~18, pp. 1--43, Apr. 2018.

\bibitem{Nesterov2004}
Y.~Nesterov, \emph{Introductory Lectures on Convex Optimization: A Basic
  Course}.\hskip 1em plus 0.5em minus 0.4em\relax Kluwer Academic Publishers,
  2004.

\end{thebibliography}
% Generated by IEEEtran.bst, version: 1.14 (2015/08/26)

\end{document}